\documentclass[useAMS,usenatbib]{mn2e}

\usepackage{graphicx} 
\usepackage{amssymb}

\newcommand {\hi} {{\rm H}\,{\small\rm I}}
\newcommand {\hii} {{\rm H}\,{\small\rm II}}
\newcommand {\kms} {\,{\rm km\,s}^{-1}}
\newcommand {\pc} {\,{\rm pc}}
\newcommand {\kpc} {\,{\rm kpc}}
\newcommand {\kmskpc} {\,{\rm km\,s}^{-1}\,{\rm kpc}^{-1}}
\newcommand {\de}{^{\circ}}
\newcommand {\mo}{\,{M}_\odot}

\newcommand{\Myr}{\,{\rm Myr}}
\newcommand{\Gyr}{\,{\rm Gyr}}
\newcommand{\K}{\,{\rm K}}
\newcommand {\moyr}{\,{M_\odot\,\rm yr}^{-1}}
\newcommand{\gsim}{\lower.7ex\hbox{$\;\stackrel{\textstyle>}{\sim}\;$}}
\newcommand{\lsim}{\lower.7ex\hbox{$\;\stackrel{\textstyle<}{\sim}\;$}}

\title[Supernova-driven gas accretion in the Milky Way]
{Supernova-driven gas accretion in the Milky Way}

\author[A. Marasco, F. Fraternali, \& J.\ J. Binney]
{A. Marasco$^{1}$\thanks{E-mail:antonino.marasco2@unibo.it},
F. Fraternali$^{1, 2}$\thanks{E-mail:filippo.fraternali@unibo.it}, 
and J.\ J. Binney$^{3}$
\\
$^{1}$Department of Astronomy, University of Bologna, via Ranzani 1, 40127, Bologna, Italy\\
$^{2}$Kapteyn Astronomical Institute,  Postbus 800, 9700 AV, Groningen, The Netherlands\\
$^{3}$Rudolf Peierls Centre for Theoretical Physics, 1 Keble Road, Oxford, OX1 3NP, UK
}

\begin{document}

\date{Accepted xxx, Received xxx}

\pagerange{\pageref{firstpage}--\pageref{lastpage}} \pubyear{}

\maketitle

\label{firstpage}

\begin{abstract}
We use a model of the Galactic fountain to simulate the neutral-hydrogen
emission of the Milky Way Galaxy. The model was developed to account for data
on external galaxies with sensitive \hi\ data.  For appropriate parameter
values, the model reproduces well the \hi\ emission observed at Intermediate
Velocities. The optimal parameters imply that cool gas is ionised as
it is blasted out of the disc, but becomes neutral when its vertical velocity
has been reduced by $\sim30$ per cent. The parameters also imply that cooling
of coronal gas in the wakes of fountain clouds transfers gas from the
virial-temperature corona to the disc at $\sim2 \moyr$. This rate
agrees, to within the uncertainties with the accretion rate required to
sustain the Galaxy's star formation without depleting the supply of
interstellar gas. We predict the radial profile of accretion, which is an
important input for models of Galactic chemical evolution.  The parameter
values required for the model to fit the Galaxy's \hi\ data are in excellent
agreement with values estimated from external galaxies and hydrodynamical
studies of cloud-corona interaction. Our model does not reproduce the
observed \hi\ emission at High Velocities, consistent with High Velocity
Clouds being extragalactic in origin. If our model is correct, the structure
of the Galaxy's outer \hi\ disc differs materially from that used previously
to infer the distribution of dark matter on the Galaxy's outskirts.
\end{abstract}


\section{Introduction}

Independent estimates of the baryon content of the Universe are provided by
the theory of primordial nucleosynthesis \citep[e.g.][]{Pagel97} and studies
of the inhomogeneity of the cosmic microwave background (CMB)
\citep[e.g.][]{Spergel+07}. The resulting baryon density proves to exceed by
a factor $\sim10$ that accounted for by the stars and observed interstellar
medium (ISM) of galaxies \citep{ProchaskaTumlinson09}. This dissonance
suggests that most baryons are in intergalactic space.  A fraction around
30\% of these baryons may comprise ionised gas associated with the $local$
Ly$\alpha$ forest \citep{Penton04}.  The rest is still missing, and probably
comprises the warm-hot medium (WHIM) that is believed to permeate most of
intergalactic space.  In this picture, spiral galaxies, like more massive
structures such as galaxy clusters, should be embedded in massive coronae
of gas at the virial-temperature. These coronae should typically extend to a
few hundred kiloparsecs from galaxy centres \citep{FukugitaPeebles06}.  The
observational quest for this medium is ongoing \citep{Bregman07}, and there
is some debate as to whether the observational constraints are compatible
with the cosmological predictions or not \citep{Rasmussen+09,
AndersonBregman10, AndersonBregman11}.

Cosmological coronae constitute a virtually infinite source of gas to feed
star formation in galaxy discs.  In fact several lines of evidence
indicate that gas accretion from the intergalactic medium (IGM) plays an
important role in galaxy evolution \citep{Sancisi08}.  Studies of the stellar
content of the Milky Way's disc show that the Star Formation Rate (SFR) in
the solar neighborhood has declined by a factor of only $2-3$ over the past
10 Gyr \citep{Twarog80, Rocha-Pinto+00, Cignoni+08, AumerB09}.  The slowness
of the decline in the SFR suggests that the Galaxy's meagre stock of cold gas
is constantly replenished. This suggestion is reinforced by the scarcity of
metal-poor G dwarfs, which arises naturally if the Galaxy constantly accretes
metal-poor gas \citep{PagelPatchett,Chiappini+97}.  Finally, studies of the
evolution of the cosmic SFR and thus the rate of gas consumption across the
Hubble time show that at any time galaxies must have accreted gas at a rate
close to their SFR \citep{Hopkins+08, Bauermeister+10}.

While it is widely agreed that the WHIM contains the bulk of the baryons, and
that disc galaxies sustain their star formation by accreting from the WHIM,
there is no consensus at to how they accrete gas. Some authors have argued that
cooling occurs spontaneously within WHIM coronae as a consequence of thermal
instability \citep{MallerBullock04}, producing cold clouds similar to the
Galactic high velocity clouds \citep{WakkerVanWoerden97}.  However, linear
perturbation analysis shows that in a corona stratified in a galactic potential,
buoyancy can very efficiently suppress these instabilities \citep{Malagoli87,
Balbus89}.  Recently, \citet{Binney+09} employed Malagoli et al.'s method to
study the growth of thermal instabilities in cosmological coronae typical of
spiral galaxies with different masses.  They found that in each case, the
combination of buoyancy and thermal conduction totally suppresses the growth
of thermal perturbations. \citet{Nipoti10} studied the growth of such
perturbations in a more realistic rotating corona, and reached similar
conclusions. 
These results were confirmed by \citet{Joung+11} using hydrodynamical
simulations based on adaptive mesh 
refinement.
They are also broadly consistent with the findings of the numerical simulations presented by \citet{Kaufmann09}, which show that a corona must have a
peculiarly flat profile of specific entropy to produce cold clouds.
However, Kaufmann et al.'s simulations did not include
thermal conduction, and \citet{Binney+09} found that conduction suppresses
perturbations in flat-entropy coronae. 
It appears therefore rather unlikely
that instabilities can grow anywhere in WHIM coronae to form clouds that will
eventually feed the star formation in the disc.

If coronae were thermally unstable, one would expect to find them studded
with clouds of cool gas. Searches of
ever-increasing sensitivity have been undertaken for \hi\ clouds around
external galaxies. These searches have consistently failed to discover
massive \hi\ clouds at large distances from galaxies, even in galaxy groups
\citep{LoSargent,Pisano+07, Irwin+09}. Deep surveys in \hi\ emission such as HIPASS and ALFALFA, indicate that massive bodies of \hi\ are always associated with
stars and thus constitute the ISM of a galaxy \citep{Doyle+05, Saintonge+08}.

What sensitive observations of nearby galaxies have revealed is that
star-forming galaxies hold significant quantities of \hi\ one or more
kiloparsecs above their discs \citep{vanDerHulstSancisi88, Fraternali+02,
Oosterloo+07}. This gas rotates around the centre of the galaxy but
at lower speed than the gas in the disc, and there are
indications that it has a net component of velocity inwards
\citep{Fraternali+02, Barbieri+05}. 

Since the temperature of \hi\ has to be lower than the virial temperature by
at least a factor 100, this \hi\ cannot be kept off the disc by pressure
forces, but must be moving essentially ballistically. Numerical simulations
of the impact of supernovae on galactic discs predict that most of the gas
ejected from the disc by a superbubble should indeed be much cooler than the virial
temperature \citep{Melioli+08}.  Observationally, there are strong
indications that in both the Milky Way and external galaxies \hi\ has been
blasted out of the disc by supernova-powered superbubbles -- prominent
extraplanar \hi\ clouds occur near regions of the disc in which the surface
density of \hi\ is depressed and there are massive young stars \citep{Boomsma+08, Pidopryhora+07}.
Another indication that star formation gives rise to extraplanar \hi\ is the
observation that galaxies such as NGC 891 with unusually large SFRs also hold
anomalously large fractions of their \hi\ as extraplanar gas.  Consequently,
there is strong evidence that extraplanar \hi\ is gas that has been blasted
out of the disc by supernova-powered superbubbles, and is orbiting in the
galaxy's gravitational field. That is, extraplanar \hi\ constitutes a
galactic fountain.

\citet[][hereafter FB06, FB08]{FB06, FB08} modelled galactic fountains with
and without accretion from the surrounding environment, and applied their
models to NGC\,891 and NGC\,2403.  They found that the kinematics of the
extra-planar gas -- rotational lag with respect to the disc and global inflow
-- could only be explained once accretion from the corona was included.
Moreover, the accretion rate required to explain the observed kinematics turned
out to be very close to the SFR in these galaxies. An aspect of this work
that did not make sense physically was the assumption that as a cloud of \hi\
moved through the virial-temperature corona, its mass had to increase as a
result of accreting coronal gas. Since the cooling time of the corona greatly
exceeds the time required for each parcel of coronal gas to flow past the
cloud, accretion should not take place.

\citet{Marinacci+10} investigated the interaction between the \hi\ clouds of
the fountain and the ambient corona using 2D hydrodynamical simulations.
They found that as the cloud moves, gas is stripped off it and mixed with the
much hotter coronal gas.  If the metallicity and pressure of the mixed gas is
high enough, radiative cooling is fast and the mixture cools to $\sim10^4\K$
within a dynamical time. Then the gradually eroding \hi\ cloud trails a wake
of cold gas behind it, and the total mass of \hi\ increases with time, just
as the models of FB08 required.  \citet{Marinacci+10} suggested that this is
the mechanism by which gas is transferred from the WHIM to star-forming
discs.

Our location within the disc of the Milky Way affords us a radically
different view of the local disc/corona interaction from views we have of
such interactions in external galaxies.  Extraplanar gas was first
observed around the Milky Way in the form of High Velocity Clouds (HVCs)
\citep[e.g.][]{Oort66}. For decades after the discovery of HVCs, their
distances and hence masses remained controversial
\citep[e.g.][]{BlitzSpergel,BraunBurton}. In the last decade it has been
established that the original HVCs are either associated with other galaxies
(the Magellanic Clouds or M31) or they lie within about $10\,$kpc from the
Galactic disc and consequently have small masses \citep{Wakker+07,
Wakker+08}. However, the HVCs represent only the tip of the iceberg of
extraplanar gas: most of this gas should be detected at Intermediate
Velocities, and indeed at such velocities there is abundant \hi\
emission \citep[][hereafter MF11]{MarascoFraternali11}. 
 Part of this emission constitutes the
  Intermediate Velocity Clouds (IVCs), bright complexes with disc-like
  metallicity located around $\sim2\kpc$ from the Sun, which have been
  regarded as nearby Galactic fountain clouds \citep{Wakker01}. 

MF11 defined a kinematic model of
the Galaxy's \hi\ layer and used it to simulate the \hi\ datacube of the
Leiden-Argentine-Bonn (LAB) survey of Galactic \hi\ emission
\citep{Kalberla+05}.
 This model reproduced well the Galactic emission at intermediate
  velocities, including the IVCs, which are interpreted as a local manifestation
  of a more global phenomenon.
In this paper we ask whether we can obtain
better-fitting datacubes by replacing the kinematic model of MF11 with the
dynamical model that FB06 and FB08 developed to account for observations of
external galaxies.  
In Section \ref{model} we describe the model we will be
using, which is an updated version of that described in FB08, and explain how
we evaluate the fit between a model datacube and that of the LAB survey.  In
Section \ref{pureFountain} we investigate whether the $\hi$ thick disc of the Milky Way can be produced by a pure galactic fountain.
Then in Section \ref{accretingFountain} we present evidence that this fountain interacts with
the ambient coronal gas. In Section \ref{sec:layer} we use our model to infer
the large-scale structure of the Galaxy's \hi\ halo, and argue that there is
a fundamental distinction between HVCs, which have an extragalactic origin,
and Intermediate Velocity Clouds, which are dominated by fountain gas.  In
Section \ref{discussion} we show that our model's parameter values are
remarkably consistent with results from earlier studies of (a)
small-scale hydrodynamical simulations of turbulent mixing, and (b) analysis
of data for external galaxies. Section \ref{conclusions} sums up and looks to
the future.



\section{The model}
\label{model}

We set up a galactic fountain model for the Milky Way by integrating orbits
of gas clouds in the Galactic potential.  The latter has been constructed
using the standard decomposition into a bulge, stellar and gaseous discs, and
a dark-matter halo.  The physical parameters and functional forms for the
various components have been taken from \citet[][Model II, p.~113]{GD2}.

The clouds are ejected from the disc into the halo, and we follow their
orbits until they return to the disc as described in FB06.  The probability
of ejection at speed $v$ at angle $\theta$ with respect to the vertical
direction is
\begin{equation}\label{probeject}
P(v,\theta)\propto\exp\left(-{v^2\over2h_{\rm
v}^2\cos^{2\Gamma}\theta}\right),
\end{equation}
 where $h_{\rm v}$ is a characteristic velocity and $\Gamma$ is a constant
that determines the extent to which clouds are ejected perpendicular to the
disc. Here we treat $h_{\rm v}$ as a free parameter but adopt
  $\Gamma=5$ from FB06 -- such a large value of $\Gamma$ implies
strong collimation of ejecta towards the normal to the plane by cool
gas near the plane.  The integration is performed in the $(R,z)$ plane
and then spread through the Galaxy by assuming azimuthal symmetry.
The number of clouds ejected as a function of radius is proportional
to the SFR density at that radius (see Section \ref{SFlaw}). 
Other parameters and properties of the model are as described in FB06
except for some updates to the model that are specified in the
following subsections.

\subsection{Phase-change}
\label{phase-change}

We follow the orbits of gas that leaves the disc at relatively low
temperatures ($\lsim 10^4 \K$) after being swept up in the expansion of a
superbubble that is powered by multiple supernova explosions.  The
virial-temperature gas filling the bubble is presumed to contain negligible
mass and to merge with the pre-existing coronal gas.  In the Milky Way,
emission from the \hi\ halo occurs mainly at negative line-of-sight
velocities \citep{vanWoerden+85}.  This observation suggests that the neutral
gas is seen mainly as it descends to the plane, so a significant fraction of
the ejected gas must be ionised, and indeed from their 3D kinematic models of
the galactic halo MF11 estimated that the rising clouds are $\sim60$ per cent
ionised.  If we are to compare our models with the LAB survey of neutral gas,
we have to find a way to estimate the fraction of a cloud that is ionised at
each part of its orbit.  

FB06 built models with and without phase-change, the latter having the whole
orbit ``visible'' as \hi\ gas, the former only the descending part.  Here we
refine this treatment as follows.  We assume that a cloud ejected from the
disc with a kick velocity $v_{\rm kick}$ will become visible (i.e.\ neutral)
only when 
\begin{equation}\label{ionfrac}
v_z(t)\!<\!v_{z,0}(1-f_{\rm ion}),
\end{equation}
 where $v_z$ is the vertical component of the cloud's velocity, $v_{z,0}\!=\!v_{\rm kick}\times\cos\theta$ and $0\!\le\!f_{\rm
ion}\!\le\!1$ is a free parameter that regulates the ``visibility'' of the
cloud.  If $f_{\rm ion}\!=\!1$, the fountain cloud is neutral only for
negative values of $v_z$ (i.e.\ in the descending part of the orbit), while
if $f_{\rm ion}=0$, the cloud will always be visible. We infer the correct
value of $f_{\rm ion}$ from the data cube (see Sections \ref{pureFountain} and \ref{accretingFountain}).

\subsection{Star-Formation Law}
\label{SFlaw}

We assume that the strength of the supernova feedback, i.e.\ the number of
gas clouds ejected per surface element, is proportional to the local SFR.
Given that our model is axisymmetric, we require an estimate of the SFR as a
function of $R$.  In FB06 and FB08 that was estimated from the surface
density of neutral plus molecular gas using the Schmidt-Kennicutt law
\citep{Schmidt59, Kennicutt98}.  The SFR was then set to zero beyond a
certain cut-off radius.  Here, we refine this procedure as follows.  We use a
sample of galaxies with known radial trends of gas surface density and SFR
density and derive a star-formation law for molecular gas.  We then use this
law to obtain the  Milky Way's SFR as a function of $R$  from the
axisymmetrised surface density of molecular gas given in \cite{bm98}. 

\begin{figure}
\begin{center}
\includegraphics[width=0.48\textwidth]{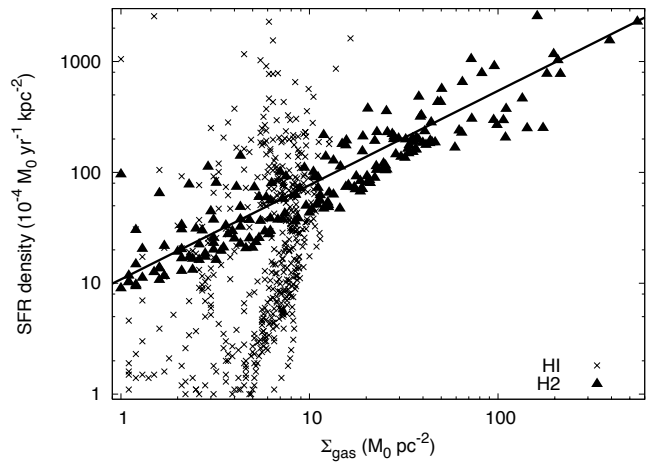}
\caption{Star formation rate densities vs gas surface densities, both
  for molecular gas (filled triangles) and neutral atomic gas
  (crosses); from the sample of nearby galaxies studied by
  \citet{Leroy+08}. 
  The points come from an azimuthal averages at different radii over the
galactic discs of $17$ galaxies. 
The curve shows the power law fit described in the text.
}
\label{fSFlaw}
\end{center}
\end{figure}

Fig.~\ref{fSFlaw} is a plot of SFR density versus the molecular gas density.
The filled triangles show data points for galaxies in the sample of
\citet{Leroy+08} that have stellar masses above $10^9 M_{\odot}$. Each point
refers to an azimuthal average at the given radius.
The scatter in the relation between
gas density and SFR is remarkably small
considering that the points come from very different galaxies.  We fitted
these points with a power law of the form:
\begin{equation}\label{eq:givesSD}
 \Sigma_{\rm SFR} = A \left({\Sigma_{\rm H_2}\over\mo\pc^{-2}}\right)^N,
\end{equation}
 where $\Sigma_{\rm SFR}$ and $\Sigma_{\rm H_2}$ are,
respectively, the SFR and the molecular gas density.  We find
\begin{eqnarray}
A&=&10.8 \pm 2.3 \times 10^{-4} \moyr\,{\rm
kpc}^{-2}\nonumber\\
N&=&0.85 \pm 0.04\ .
\end{eqnarray}
 The slope of this relation is very different from the standard
Schmidt-Kennicutt law, mainly because we are considering only molecular gas
\citep[see also][]{Krumholz+11}.  Including \hi\ would increase the slope to
about $1.3-1.4$, however the average \hi\ surface density correlates very
little with the SFR density -- see the grey crosses in Fig.~\ref{fSFlaw} and
\citet{Kennicutt+07}.  
Therefore, given that the distribution of 
molecular gas in the disc of the Milky Way is known reliably, we prefer to
obtain the SFR density from equation (\ref{eq:givesSD}). Our
results do not depend strongly on the SF law.
 
\begin{figure*}
\begin{center}
\includegraphics[width=\textwidth]{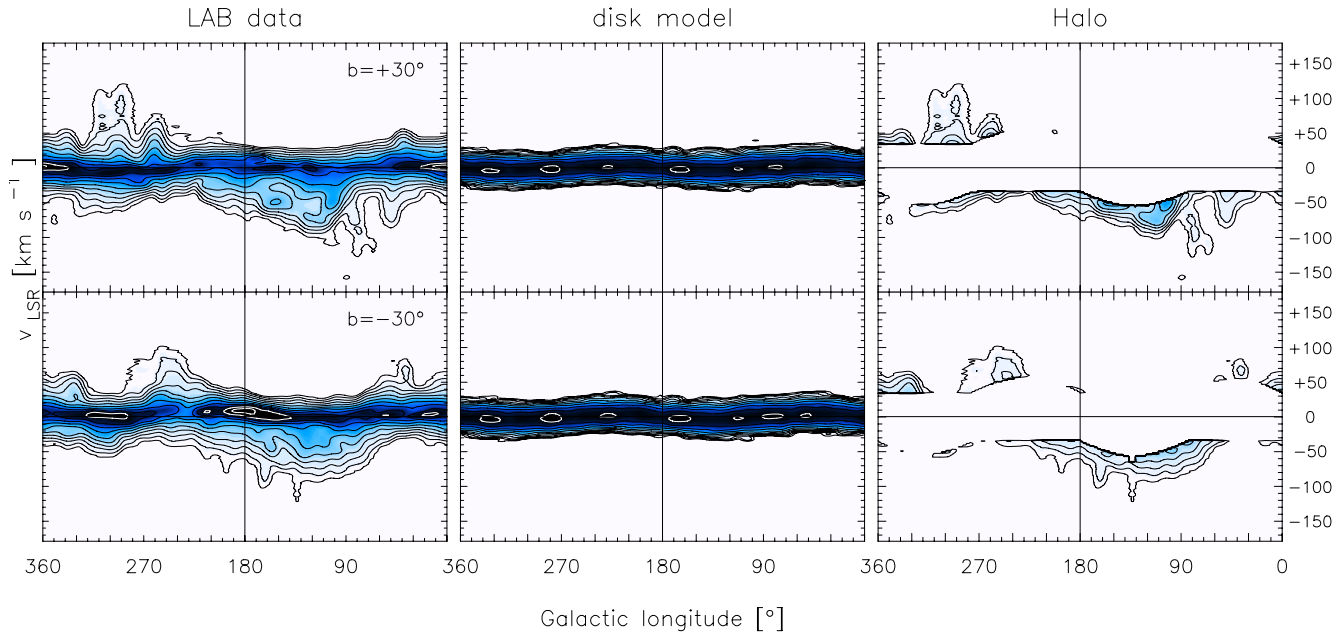}
\caption{
Longitude-velocity diagrams at latitudes $\pm 30\de$ above and below the 
mid-plane of the Galaxy.
 \emph{Left panels}: LAB survey.  \emph{Middle panel}: disc model used
 throughout this paper. 
 \emph{Right panel}: LAB survey after the region of the disc has been
 removed, the remaining emission is used to compared our models with
 the data.
Each datacube is smoothed to $8\de$ resolution. 
Contour levels in brightness temperature range from $0.04\K$ to
$81.92\K$ scaling by a factor $2$.
}
\label{fData}
\end{center}
\end{figure*}

\subsection{Supernova-driven gas accretion}
\label{SN-driven}

The pure galactic fountain model described above is modified to include
interaction with the coronal gas according to FB08.  We assume that the main
mechanism at work is that described by \citet{Marinacci+10}: the
Kelvin-Helmholtz instability generates a turbulent wake behind each cloud in
which material stripped from the cloud mixes with coronal gas, enhancing the
metallicity and density of the latter and substantially decreasing its
cooling time.  As a consequence, part of the corona condenses into the wake
and the mass of ``cold'' gas increases.  The combined mass of cold gas in the
cloud and the wake grows with distance along the cloud's path through the
halo very much as proposed by FB08.  We demonstrate this quantitatively in Section
\ref{compSim}.

Condensation in the wake modifies the kinematics of the cold gas in a way
that depends on the kinematics of the corona. The latter is not known a
priori, but must be strongly influenced by this interaction because the
portion of the corona that interacts with fountain clouds does not contain
much mass, and the interaction has been taking place throughout the disc's
lifetime.  FB08 considered the effect of the drag between the clouds and the
corona and concluded that, if there were no exchange of gas between the
clouds and the corona, ram pressure between the cloud and the corona would
force the corona to corotation with the cold gas in a time shorter than the
fountain's dynamical time. In reality much of the coronal gas that acquires
momentum from a fountain cloud subsequently condenses in the cloud's wake, so
the momentum that is transferred to the coronal gas is in the end not lost by
the whole body of cool gas. \cite{Marinacci+11} studied the transfer of
momentum between fountain clouds and the corona in hydrodynamical simulations
and found that there is a net transfer of momentum to the corona until the
speed at which the clouds move through the corona falls to $50-100 \kms$,
depending on the physical properties of the system.  Below this threshold
speed, net momentum transfer to the corona ceases as a result of coronal gas
cooling in the cloud's wake. In view of the short timescale on which
ram-pressure can change the rotation rate of the corona, it is natural to
assume that the corona has reached the velocity at which all the angular
momentum imparted by ram pressure is subsequently recovered through
condensation of coronal gas in the wake. Thus we assume that the rotation
velocity of the corona is
 \begin{equation}\label{vphiambient}
v_\phi(R,z) = \sqrt{R\frac{\partial\Phi(R,z)}{\partial R}} - v_{\rm lag},
\end{equation}
 where $\Phi(R,z)$ is the potential of the Galaxy and $v_{\rm lag}=75\kms$ is
the (constant) velocity offset between clouds and corona at which condensation
recaptures angular momentum transferred by ram pressure.  

In this picture, the mass of cold gas (cloud$+$wake) associated with any
cloud increases as 
 \begin{equation}\label{eq:mdot}
\dot m = \alpha m,
\end{equation}
 where $\alpha$ is the condensation rate, while the combined effect of condensation and drag decreases the velocity centroid of the
cold gas at a rate (FB08)
 \begin{equation}\label{eq:vdot}
\dot v=-\left(\alpha+\frac{1}{t_{\rm drag}\left(1+t/t_{\rm drag}\right)}\right)(v-v_\phi)
\end{equation}
 where $t_{\rm drag}$ is the time at which the relative velocity
 between the cold gas and the corona halves due to the drag only. 
Since $t_{\rm drag} \! \propto \! |v-v_\phi|^{-1} \! \sim \! v_{\rm lag}^{-1}$, the drag efficiency decreases as the kinematics of the corona approaches that of the disc.
 We assumed a constant value for $t_{\rm drag}$ of $800\Myr$, according 
 to the results of \citet{Marinacci+11} for $|v-v_\phi| = 75\kms$ (see Section \ref{compSim}), and we fitted the parameter $\alpha$. 
As we see below (Section \ref{accretingFountain}), in order to reproduce the data we need $\alpha\!>\!1/t_{\rm drag}$, i.e.\ the condensation is
the dominant process at work, while the drag plays only a secondary
role.
Note that fixing $t_{\rm drag}$ at $800\Myr$ implies that we are considering clouds with masses and sizes similar to those of Marinacci et al.'s simulations ($\sim10^4 \mo$,\,$100\pc$).

\subsection{Comparison with the data}
\label{comparison}

The comparison between the models and the data is performed by building model
datacubes that resemble the LAB datacube (see FB06 and MF11).  After each
timestep of the integration, the positions and velocities of the particles
are projected along the line of sight of an observer located in the mid-plane
of the Galaxy, at a radius $R_0 = 8.5 \kpc$, and moving at a circular speed
$v_{\rm LSR}= 220 \kms$.  Using the same gas density for each cloud, we
construct from these positions and velocities an artificial datacube. The
artificial datacube is smoothed to $8\de$ resolution and then compared with the LAB
datacube at the same resolution.  The comparison is performed only in the
portion of $(l,b,v_{\rm LOS})$ space to which the thin disc should
not contribute.  Following \cite{Wakker91}, we use the \emph{deviation velocity}
$v_{\rm DEV}$ to identify this region.  We model the footprint of the thin
disc as described in MF11 and exclude data with $|v_{\rm DEV}| < 35 \kms$.
We also excluded the \hi\ emission of the Magellanic Clouds and Stream, the
Stream's Leading Arm, the GCP complex, the Outer Arm and all external
galaxies. In the following we refer to the surviving part of $(l,b,v_{\rm
LOS})$ space as the \emph{halo region}.  The total model flux in the halo
region is normalized to the corresponding LAB flux. This step fixes the mass
of the \hi\ halo.

Although the quantitative comparison between models and data is done only in
the halo region, in all the plots of this paper a thin disc has been added
for presentation purposes.  This disc has a density profile taken from
\citet{bm98} and a scale-height taken from \citet{Kalberla+07}. The velocity
dispersion decreases linearly from $12\kms$ in the Galactic centre to $6\kms$
at $R_0$, and it remains constant at larger radii.  

The leftmost column of Fig.~\ref{fData} shows two longitude-velocity ($l,v$)
plots at latitudes $\pm 30\de$ extracted from the LAB survey.  The middle
column of the same figure shows our model disc and the rightmost column shows
the emission in the halo region of the LAB datacube. All the plots are centered at the anti-centre ($l\!=\!180\de)$. Note that our cut is
rather conservative and there should be no contamination from disc emission
in the halo region.

In the application of the model to external galaxies, the fountain clouds
have been modelled as point particles because individual clouds, with typical
sizes $\sim100\pc$, were unresolved in the data (FB06).  In the Milky Way
nearby clouds and their wakes will be resolved, so we have to check at each
time step whether the combined emission of a cloud and its wake is resolved
at the angular resolution of the data.  If the emission is resolved, we
spread the flux in a circular region around its centre. The flux density is
assumed to decrease Gaussianly with distance from the centre, with FWHM equal
to the apparent size of the cloud and its wake.  Given that the data are used
at $8\de$ resolution, a cloud and its wake are resolved only if their
distance is smaller than $700 \times \frac{D_{\rm cl}}{100\,{\rm pc}} \pc$,
where $D_{\rm cl}$ is the system's lengthscale.  According to the simulations
of \citet{Marinacci+10}, turbulent wakes have sizes of $1-2 \kpc$, so they
will be resolved out to distances of about $10 \kpc$.  The pattern of
emission from a cloud and its wake is quite elongated, having axis ratio
$\sim0.5$, but the long axis of the emission is randomly orientated, so our
circular Gaussian smoothing should provide an adequate representation of the
total emission from an ensemble of clouds.  We tested the effect of
resolution by increasing $D_{\rm cl}$ up to $2 \kpc$ and found that the model datacube
looked just like a smoothed version of that produced with smaller values of
$D_{\rm cl}$.  In the following we use models with small clouds, $D_{\rm cl}=200
\pc$, for which the datacube can be computed significantly more quickly.

Once we have a model datacubes at the same resolutions and with the same
total flux as the LAB datacube, we compare them quantitatively by calculating
the residuals between them.  This calculation is performed by adding up the
differences between the models and the data, pixel by pixel, in the halo
region.  We used absolute differences, squared differences and also weighted
residuals, i.e.\ $|$data$-$model$|/$(data$+$model).  The residuals are calculated for a
set of models with different input parameters such as the kick velocities
$h_{\rm v}$ of the fountain clouds, and the model with the smallest residuals
is our ``best model''.  In the following section we describe our best pure
fountain model and our best fountain $+$ condensation model, showing that the
latter provides a better description of the Galaxy's \hi\ data.

\section{Results}

\subsection{A pure galactic fountain}
\label{pureFountain}

\begin{figure}
\includegraphics[width=0.5\textwidth]{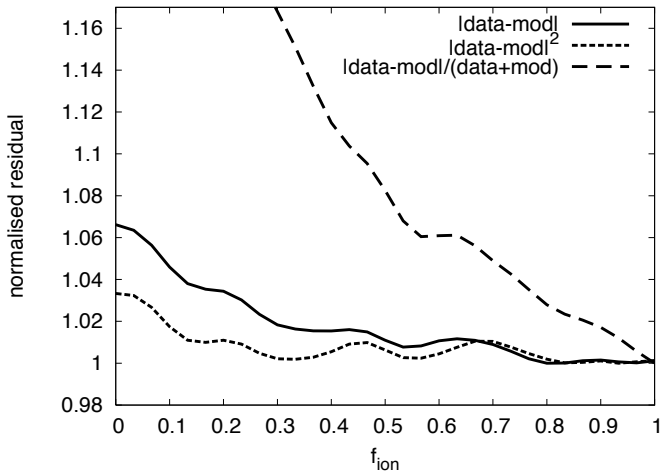}
\caption{
Residuals between models and data as a function of $f_{\rm ion}$ for the pure galactic fountain with $h_{\rm v}=70\kms$, evaluated using absolute differences (\emph{solid line}), squared differences (\emph{short-dashed line}) and weighted differences (\emph{long-dashed lines}). All residuals have been divided by their respective minimum.}
\label{res_pure}
\end{figure}

In a pure galactic fountain, the angular momentum of clouds is constant and
orbits depend only on the characteristic kick velocity $h_{\rm v}$ of the
clouds. However, the shape of the model datacube is also affected by the
ionised fraction $f_{\rm ion}$, which has to be
treated as a freely variable parameter.

\begin{figure*}
\begin{center}
\includegraphics[width=\textwidth]{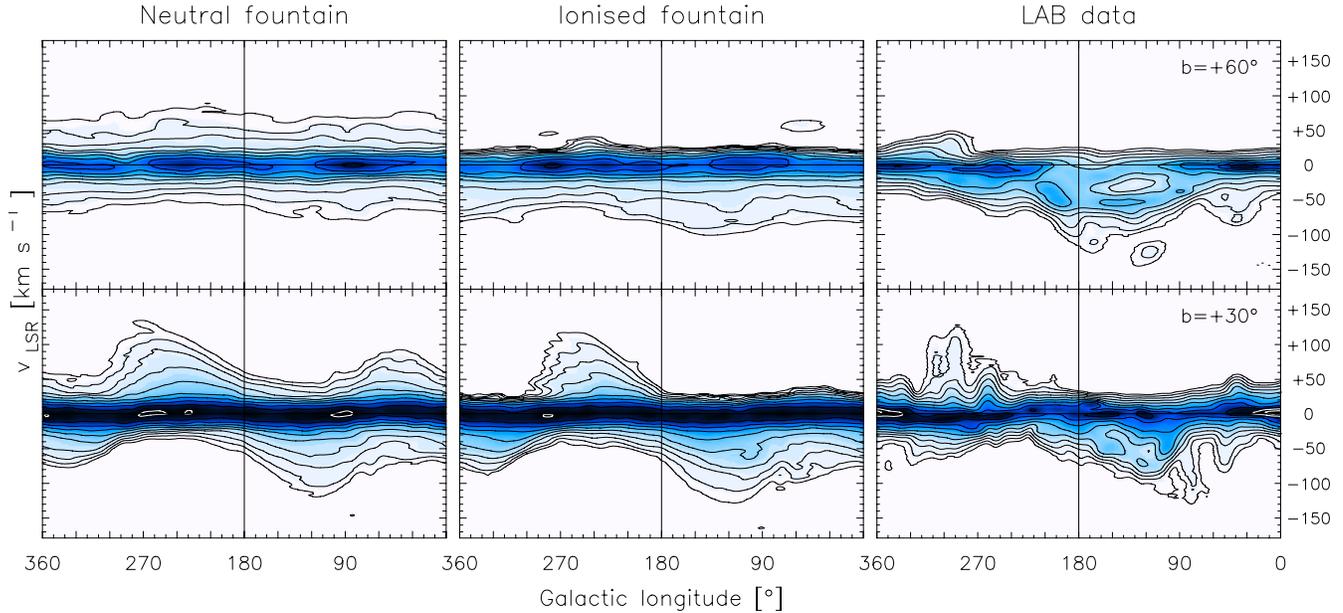}
\caption{Longitude-velocity ($l,v$) diagrams at $b\!=\!60\de$ (top
  panels) and $b\!=\!30\de$ (bottom panels).
\emph{First column}: pure fountain model without phase-change
($f_{\rm ion}=0$); \emph{second column}: best pure fountain model
($f_{\rm ion}=1$); \emph{third column}: the LAB data.
Each datacube is smoothed to $8\de$ resolution. 
Contour levels in brightness temperature range from $0.04\K$ to
$81.92\K$ scaling by a factor $2$. 
}
\label{fPureFountain}
\end{center}
\end{figure*}

We have calculated the residuals for $h_{\rm v}$ in the range $(30,120)\kms$ and $f_{\rm ion}$ between $0$ and $1$.  The weighted and non-weighted
residuals give different results but the confidence contours overlap for
$h_{\rm v} \simeq 70 \kms$ and $f_{\rm ion}=1$.  
In Fig.~\ref{res_pure} we show the normalized residuals as a function of $f_{\rm ion}$ for $h_{\rm v}\!=\!70\kms$.
Clearly the differences between models and data decrease as $f_{\rm ion}$ approaches $1$, regardless of the method of comparison.
The mass of the \hi\ halo derived for this model is
  $M_{\rm halo}=5.2\times10^8 M_{\odot}$.  
Here and below the halo masses are estimated by excluding fountain
clouds at $|z|\!<\!0.4\kpc$, which is roughly the blowout height of a 
superbubble \citep[see][]{Spitoni+08}.
Taking into account also clouds at heights $0.2\!<\!|z|\!<\!0.4\kpc$,
the \hi\ halo mass would become $M_{\rm halo}=7.2\times10^8
M_{\odot}$.

Fig.~\ref{fPureFountain} shows two representative $l,v$ diagrams (centred on
the anticentre) at latitudes $b\!=\!60\de$ and $b\!=\!30\de$ for: the best
pure galactic fountain model (middle column); the same model without
phase-change (left column); the data (right column).  The difference between
the two models is striking, especially at higher latitudes.  The $l,v$ plots
for the neutral ($f_{\rm ion}=0$) fountain are rather symmetric with respect
to the zero velocity line.  In the ionised fountain by contrast, the gas
appears systematically located at negative velocities.  This effect is a
consequence of clouds of the ionised fountain being visible only as they fall
back to the disc.  The data clearly display this preference for more negative
than positive velocities. 

The leftmost column in Fig.~\ref{fBestModels} shows additional $l,v$ diagrams
for the best pure fountain model. It reproduces the most prominent features,
so it gives a good description of the data.  It is interesting to compare
Fig.~\ref{fBestModels} with Fig.~6 of MF11, which shows the prediction of an
optimised kinematic model.  In several locations the dynamical model fits the
data better than the best kinematic model of MF11.\footnote{Note that the
plots in MF11 are centred at the galactic centre.} For example, at latitudes
$b=\pm15\de$ the superiority of the present model is very clear in the
central regions of the Galaxy, $0\de<l<60\de$ and $300\de<l<360\de$. At
latitudes $b=\pm30\de$, the superiority is clear at $l\sim270\de$ ($-90\de$
in MF11).

The kinematic model was built under the assumption that the scaleheight and
kinematic parameters of the halo (rotation velocity gradient, vertical and
radial velocities) are independent of radius.  Although the assumption is
restrictive, the model has been fitted to the data by freely varying these
quantities, regardless of their dynamical plausibility.  The fact that a
dynamical model with only two free parameters fits the data better is
remarkable, and strongly supports the idea that the \hi\ halo of the Milky
Way is produced by supernova feedback from the Galactic disc.

\begin{table*}
  \centering
\begin{minipage}{140mm}
   \caption{Best galactic fountain models for the Milky Way.}\label{tModels}
   \begin{tabular}{lccccccc}
     \hline
     Fountain model    &$h_{\rm v}$&$M_{\rm HI}^{~~~ a}$ & $M_{\rm HI+HII}^{~~~~~~~~~ a}$&$f_{\rm ion}$&$\alpha$   &$M_{\rm accr}$ & Residuals$^c$\\
                                 &($\kms$)  &($\mo$) &($\mo$)&    &(Gyr$^{-1}$)&($\moyr$) & \\
     \hline
Neutral (pure)              & $70^b$& $4.6\times10^8$  &  $4.6\times10^8$ &  $0.0^b$ & $0.0^b$  &  0.0 & 1.60\\
Ionised (pure)                 & 70& $5.2\times10^8$  & $10.4\times10^8$   &  $1.0$ & $0.0^b$  &  0.0 & 1.48\\
With corona condensation & 70& $2.7\times10^8$  &  $3.0\times10^8$ & 0.3   &   6.3   &  1.6 & 1.26\\
     \hline
   \end{tabular}\\
$^a$ estimated above $400\pc$ from the plane; $^b$ fixed; $^c$ calculated as absolute differences, the closer to 1 the better the fit.\\
\end{minipage}
\end{table*}

\subsection{Including the condensation of the corona}
\label{accretingFountain}

\begin{figure*}
\includegraphics[width=\textwidth]{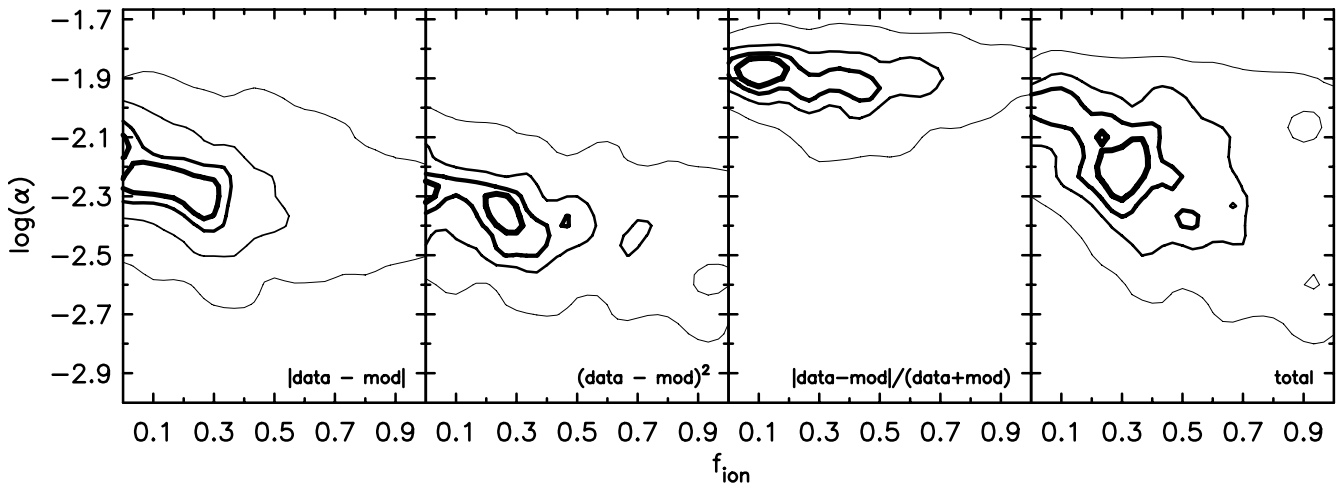}
\caption{Contour plots of residuals between models with corona condensation and data evaluated in the parameter space ($\log(\alpha)$,$f_{\rm ion}$) for $h_{\rm v}=70\kms$. Different panels compare models and data in different ways, as labelled on bottom. In each panel all values have been divided by the respective minimum. The fourth panel is the sum of the previous three, divided by the resulting minimum. Contour levels at $1.005, 1.01, 1.02, 1.05$.
}
\label{residuals}
\end{figure*}

\begin{figure*}
\begin{center}
\includegraphics[width=\textwidth]{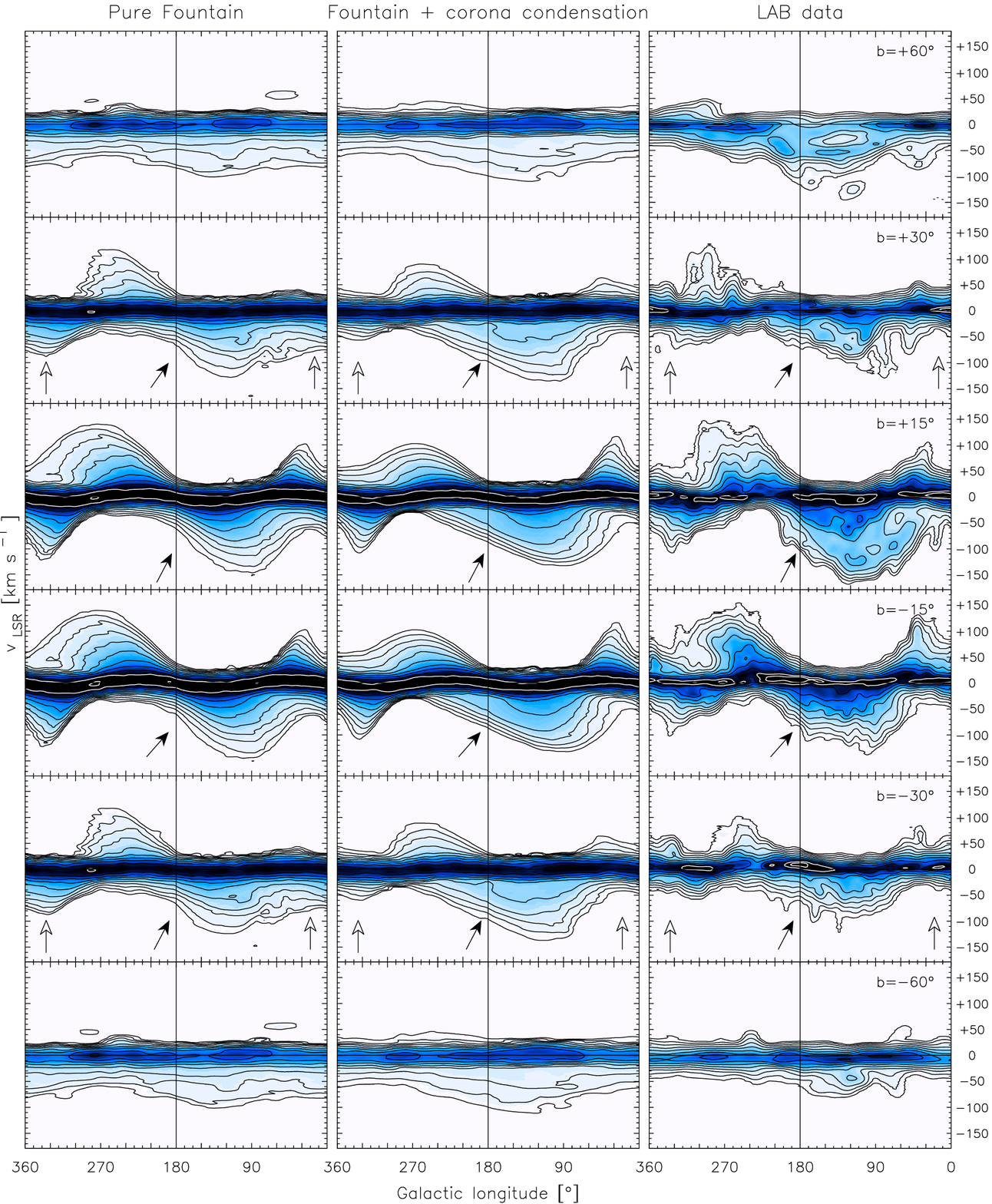}
\caption{
$l,v$ slices at 6 different latitudes indicated at the
  top right corner of the rightmost plots.
\emph{First column}: best pure Galactic fountain;
\emph{second column}: best fountain with condensation of the corona; 
\emph{third column}: the LAB data. 
Each datacube is smoothed to $8\de$ resolution. Contour levels in
brightness temperature range from $0.04\K$ to $81.92\K$ scaling by a
factor $2$.
The arrows show regions where radial motions are present and the model
with coronal condensation reproduces the data better than the pure fountain model,
see text. 
}
\label{fBestModels}
\end{center}
\end{figure*}
 
The best pure fountain model described above is extreme in that it requires
the full ionisation of ejected clouds and a large mass of the gaseous halo (see Table \ref{tModels}).  
Moreover there are several locations in the
datacube where the model systematically fails to reproduce the data in
detail.  The arrows in Fig.~\ref{fBestModels} indicate such regions. In these
regions radial motions are important.  For instance filled arrows show the
region around the anticentre, $150\de<l<250\de$, where the data, almost at
every latitude, have a broader distribution in velocity than the pure
fountain model.  Gas in this region is flowing towards the centre of the
Galaxy from beyond the Solar circle.  In a pure galactic fountain, orbits
seldom move in this way (FB06) and independently of the ionised fraction they
do not produce much emission in the marked region.  

We now show that including the interaction between clouds and the corona as
described in Section \ref{SN-driven} substantially improves this situation.
The interaction is parametrised by the specific condensation rate $\alpha$.
The larger $\alpha$ is, the faster gas condenses from the corona, the greater
the rotational lag of the \hi, and the more prominent is inward radial motion
of \hi\ (see FB08).

Now the residuals between the data and the models are minimised by varying
three parameters ($h_{\rm v},f_{\rm ion},\alpha$).  We assume that
condensation and drag are suppressed close to the Galactic plane
($|z|<0.4\kpc$). A full analysis of this three-dimensional parameter
space would be too expensive computationally, so we explore only
the space ($\alpha$,$f_{\rm ion}$) and fix the kick velocities at three
values: $h_{\rm v}=60$, $70$ and $80\kms$.  For each value of $h_{\rm v}$ we
build a map of the residuals between models and data by varying the other
parameters in the ranges $0<f_{\rm ion}<1$ and $-3<\log(\alpha/\Gyr^{-1})<-1.67$
with a step of $0.033$ in both directions.  This map is then smoothed to the
resolution of $0.1\times0.1$ in order to minimise the stochastic effects due
to the probability distribution function (eq.~\ref{probeject}) used to build
our models.  As in the case of a pure fountain, different residuals give
different results. Regardless of $\alpha$ and $f_{\rm ion}$, absolute and
squared differences have lower values for $h_{\rm v}=80\kms$, while relative
residuals favour $h_{\rm v}=60\kms$. Hence we set our kick velocity in
the middle and focus on models with $h_{\rm v}=70\kms$.

The first three panels in figure \ref{residuals} show the residual maps
obtained with $h_{\rm v}=70\kms$ for different types of residuals.  In each
panel all values are divided by the respective minimum, thus they are
dimensionless and $\ge1$.  Combining these three panels we obtain the fourth
(rightmost) panel, which is the sum of the previous three divided by the
resulting minimum, which occurs at $f_{\rm ion}=0.3$ and $\log(\alpha)=-2.2$
(so $\alpha=6.3\Gyr^{-1}$). For this best model the mass of the \hi\ halo is
$2.7\times10^8 M_{\odot}$.  Table \ref{tModels} lists the
parameters.

The middle column of Fig.~\ref{fBestModels} shows that the model with coronal
condensation reproduces the emission near the anticentre
better than the pure fountain model.  The open arrows (latitudes $b=\pm 30
\de$) indicate other regions where the inclusion of coronal condensation
improves the fit to the data.  These are regions in which \hi\ is flowing out
from the inner disc, and the pure fountain provides too much emission. 

For our best-fitting value of $\alpha$, the condensation rate of
  coronal gas into the clouds' wakes is $\dot M=1.6\moyr$. 
Correcting this value for the {\rm He} content increases the
condensation rate to $2.3\moyr$. This value is remarkably close 
to the Galaxy's SFR, which we assume to be $\sim3\moyr$
\citep{Diehl+06}, of which $\sim1\moyr$ should be accounted for by gas 
ejected from stars in the disc.  
We consider it highly significant that $\dot M$ deduced from an \hi\
survey should agree so well with that required to sustain the SFR,
which is obtained without reference to \hi.

The exact value of the condensation rate $\dot M$ depends on several factors.
An important parameter is the value of $v_{\rm lag}$ (eq.~\ref{vphiambient}),
which regulates the rotation of the corona and therefore the efficiency of both the
drag ($t_{\rm drag}\propto v_{\rm lag}^{-1}$) and the condensation \citep{Marinacci+11}. However, experiments show that the
dependence of $\dot M$ on the adopted value of $v_{\rm lag}$ is not
strong: ($v_{\rm lag}=50 \kms,\,t_{\rm drag}=1200\Myr)$  yields $\dot M=2.8\moyr$ ($3.9\moyr$ when corrected for He), while ($v_{\rm lag}=100 \kms,\,t_{\rm drag}=600\Myr$) yields $\dot M=1.0\moyr$ ($1.4\moyr$ when corrected for He).
It is interesting to note that $\dot M$ depends very little on the
criterion used to minimise the residuals.
Although the three criteria yield rather different values for the best-fitting $\alpha$ (see Fig.\ \ref{residuals}), $\dot M$ only varies between 1.5 and 1.7$\moyr$ given that the halo masses also vary between the three models.

Our parametrisation assumes that $\alpha$ is spatially constant.
Fortunately, the density of the corona should not vary too much in the
regions where the interaction with the fountain is efficient  \citep{Marinacci+11}, so
spatial variation of $\alpha$ is not expected to give significantly different
results.

On the whole, the dynamical modelling of the Galactic \hi\ halo performed in
this work is a success.  The results are consistent with the conclusions MF11
drew from their kinematic models. They found negative values for both the
average vertical and the radial velocities of the halo material, and deduced
that ejected clouds must be partially ionised (but in a fraction larger
than our current determination). Our dynamical analysis identifies these
properties as arising from the sharing of angular
momentum with gas accreted from the lagging corona. Finally, our
Galactic fountain is dynamically sustainable.  It requires a rate of
kinetic-energy injection equal to $6.2\times10^{39}{\rm erg}\,{\rm s}^{-1}$,
which with a supernova rate SNR$\,=0.03 \moyr$ and an energy per supernova of
$1\times10^{51} {\rm erg}$ corresponds to an efficiency of about 0.7\%.

\section{Properties of the Galactic \hi\ layer}\label{sec:layer}

Now that we have a model that reproduces all the main features of the LAB
datacube, we can investigate in detail the physical properties of the
Galaxy's extraplanar \hi\ layer.  

\subsection{Rotation versus height}

\begin{figure}
\begin{center}
\includegraphics[width=0.5\textwidth]{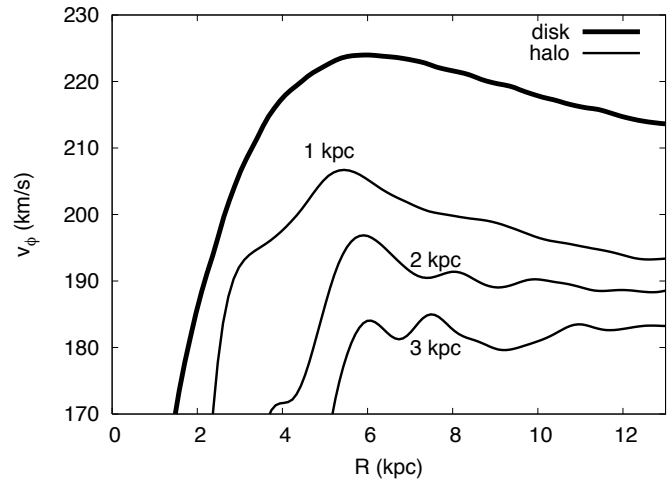}
\caption{
Rotational velocities for the \hi\ layer of the Milky Way at different
heights from the plane (thin lines), compared to the disc rotation curve (thick line).
The velocities are derived from our best mixing fountain model by
taking the weighted average of $v_{\rm \phi}$ at that $(R,z)$
location.
}
\label{rotCurve}
\end{center}
\end{figure}

Fig.~\ref{rotCurve} shows the rotation curve of the Galaxy's extra-planar gas
at different distances from the plane as a function of $R$.  These values
were derived from our best model by simply taking the weighted mean of the
azimuthal components of the velocities of the particles at different heights.
As expected, the halo gas is lagging the rotation of the disc (black curve).
Roughly 50\% of this lag is due to gravity (FB06) whilst the other half is
produced by a combination of the interaction with the corona and the
phase-change. At $R_0$ the
gradient is about $-14.3\pm1.1\kmskpc$ in excellent agreement with the average
value derived in the inner regions of the Milky Way by MF11 ($-15 \pm 4
\kmskpc$), which also coincides with the value derived for NGC\,891 by \citet{Oosterloo+07}.

\subsection{Thickness of the \hi\ layer}
\label{thickness}

\begin{figure}
\begin{center}
\includegraphics[width=0.5\textwidth]{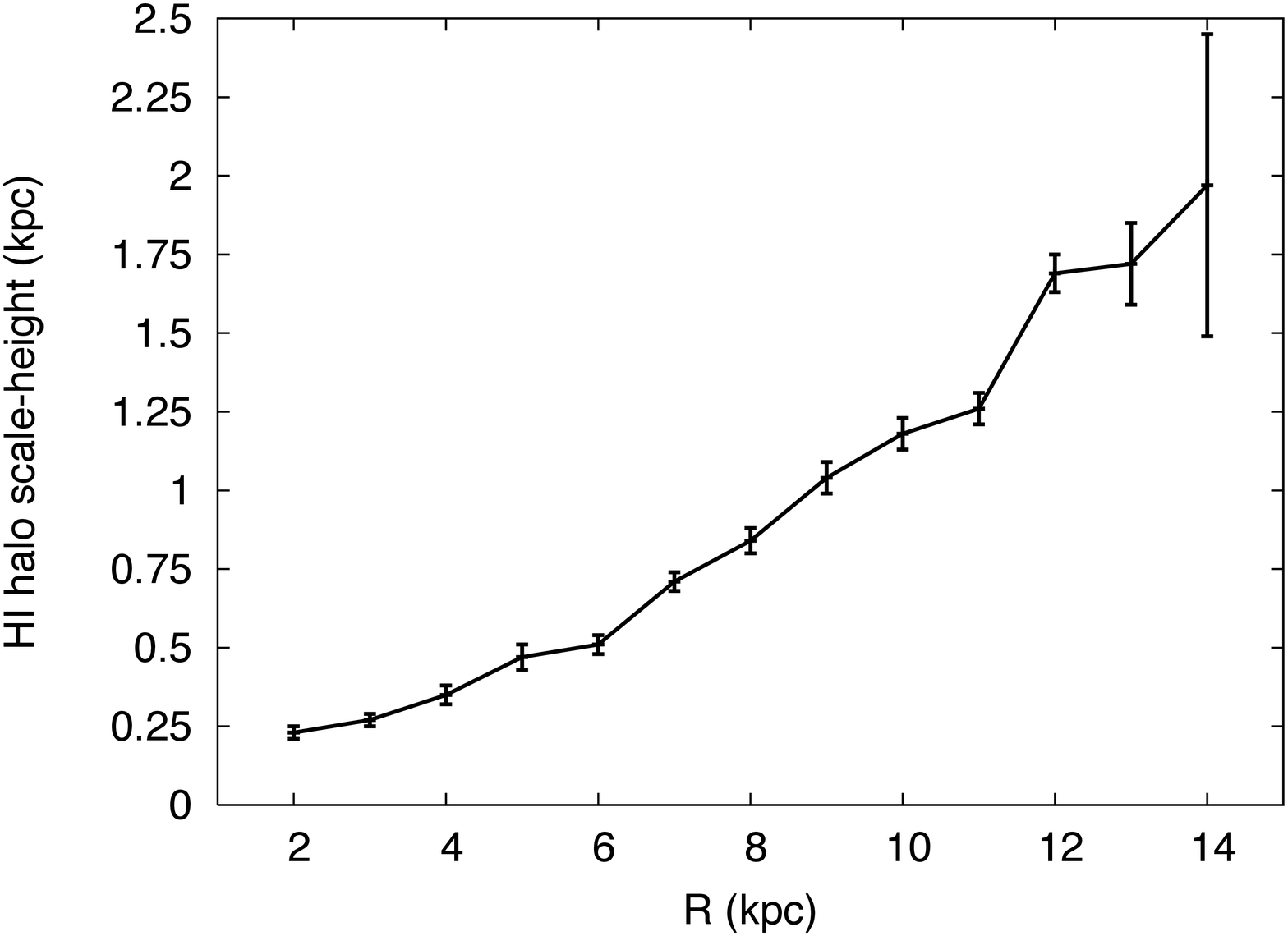}
\caption{
Scale-height as a function of $R$ for our best halo model including
condensation of the coronal gas. 
}
\label{scaleheight}
\end{center}
\end{figure}

A second fundamental property of the Galaxy's \hi\ halo is its thickness.  To
obtain it we fitted the vertical density profiles in our models at different
radii with exponential functions.  Fig.~\ref{scaleheight} shows the
(exponential) scale-height of the \hi\ halo of our best model as a function
of $R$. 
The thickness increases with $R$ because the gravitational restoring
force to the plane diminishes outwards (FB06).  The halo density decreases
abruptly for $R\!>\!14\kpc$, so its thickness cannot be reliably
determined at larger radii.
The shape of the scaleheight as a function of R is partially due to the assumption of our model that the kick velocity $h_{\rm v}$ does not change with radius (for a discussion see FB06).
In the inner Galaxy there has long been evidence for a population of \hi\ clouds extending up to $\sim1\kpc$ above the midplane \citep{Lockman84, Lockman02, Ford+10}. \citet{MarascoFraternali11} assumed a constant thickness for the halo and derived a value of $1.6\kpc$ using a $\rm{sech}^2$ formula, which
corresponds to $\sim 800\pc$ for an exponential function. This value agrees
well with the average of the scaleheights plotted in Fig.~\ref{scaleheight}.

\citet{Kalberla+07} studied the vertical structure of the Galaxy's \hi\
layer on the assumption that the layer is in hydrostatic equilibrium. Using
this assumption, they inferred that while at $R_0$ the \hi\ layer extends
only a few hundred parsecs from the plane, at $R\simeq35\kpc$ the vertical
structure of the \hi\ layer can be fitted by a Gaussian distribution with
scale height $2.5\kpc$. Thus they deduced a very extended \hi\ layer with strong flaring.
They also inferred the existence of a massive ring of dark matter between
$R=13$ and $18.5\kpc$. 
The assumption of hydrostatic equilibrium leads, however, to
models that fail to match the data correctly \citep[see][]{Barnabe06,Marinacci+10b}. 
If \citet{Kalberla+07} had used the prediction of a fountain
model for the distribution of \hi\ near $R_0$ rather than a model based on
hydrostatic equilibrium, they would have come to different conclusions
regarding the radii responsible for each quantity of emission in the \hi\
datacube.
Consequently, the distribution of the matter and the structure of the
\hi\ flare in the Milky Way should be re-derived.
While this topic merits further work, our comparisons of the
fountain model with the data yield no compelling evidence for substantial
flaring of the \hi\ layer much beyond $R\sim 15 \kpc$.  

\subsection{Accretion and circulation of gas}

\begin{figure}
\begin{center}
\includegraphics[width=0.5\textwidth]{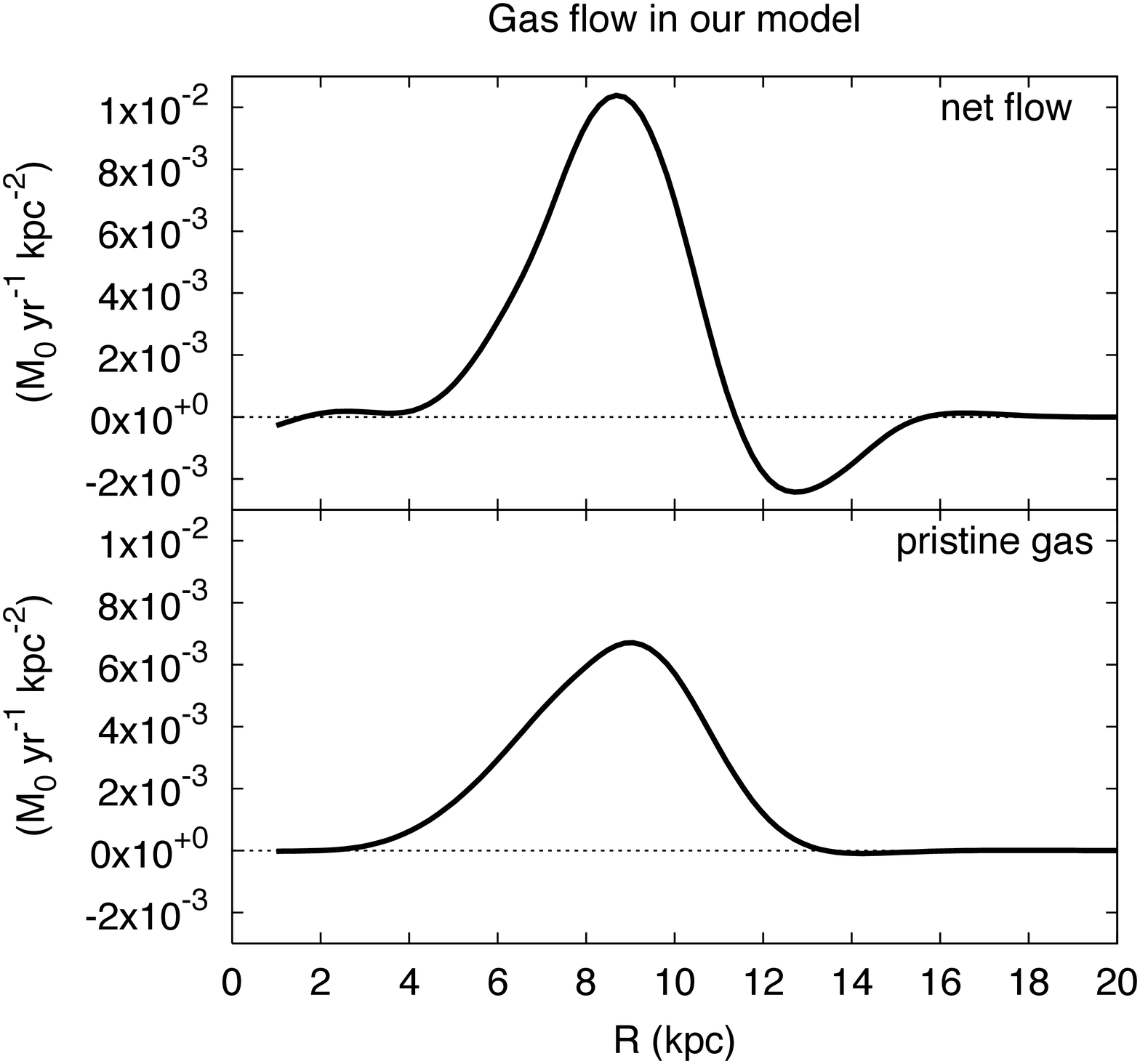}
\caption{
Hydrogen flow as function of $R$ predicted by our best model with
condensation of the corona. Inflows (outflows) are represented by positive
(negative) values. 
\emph{Top panel}: net flow ($\hbox{inflow} - \hbox{outflow}$) produced by the Galactic fountain.
\emph{Bottom panel}: accretion of pristine gas from the corona onto the disc. 
The global accretion rate is $1.6\moyr$.
}
\label{accretion}
\end{center}
\end{figure}

Infall of metal-poor gas to the star-forming disc is an essential ingredient
of current models of the Galaxy's chemical evolution
\citep[e.g.][]{Chiappini+01,SchoenrichBinney09}. The predictions of such
models depend to a significant extent on the radial profile of the infall,
but hitherto there has been no credible way of determining this profile.  The
prediction of our model for this profile is shown in Fig.~\ref{accretion} (bottom panel).
The curve shows the pristine gas that, condensing
from the corona onto the fountain cloud wakes, follows the cloud
orbits back to the disc.  
The shape of the accretion profile is due to the variation of the mass
outflow and the orbital time with radius.
At $R\la3\kpc$ the specific accretion rate essentially vanishes
because the orbits of the fountain clouds are confined within few
hundred of parsecs from the disc (see Fig.~\ref{scaleheight}).
It then rises to a peak at $R\simeq9\kpc$ as in this region both
the orbital time and the star formation are sufficiently high.
At $R>9\kpc$ the accretion rate falls again due to the low level of
star formation at these radii, and drops to about zero beyond $R=13
\kpc$. 
The integral of this curve over the disc surface gives a global
accretion rate of $\dot{M}=1.6 \moyr$.
Note that the peak in the accretion rate lies well beyond the peak in
the Milky Way's SFR, which occurs around $R\!=\!4\kpc$. 
This implies the need for a redistribution of gas in the disc, the
study of which is beyond the scope of the present paper \citep[but see for
instance][]{SchoenrichBinney09,SpitoniMatteucci11}.

The top panel of Fig.~\ref{accretion} shows the net flow (\hi+\hii) as function of $R$, obtained as the difference between the rate at which gas arrives at a given radius and the rate at which supernovae eject gas from that radius. 
The net flow profile differs from the accretion profile because fountain clouds do not fall back onto the disc at the same radius they are ejected from.
As $R$ increases, the orbits of fountain clouds start to get
an inward component because of the interaction with the corona, thus
clouds land at smaller and smaller radii. 
In particular, most of the clouds ejected at $11\!<\!R\!<\!15\kpc$ fall back onto the disc at $6\!<\!R\!<\!11\kpc$. 
As a consequence, a `draining region' forms at $11\!<\!R\!<\!15\kpc$, while inside this annulus the amount of gas
falling onto the disc significantly exceeds the outflowing counterpart: about $70\%$ of this inflow comes from the condensation of the corona,
the remaining part is due to the described gas circulation.
On the whole, the fountain circulation contributes to $76\%$ of the halo mass, while the accreted gas is $24\%$. 
This latter is similar to that derived by FB08 for NGC\,891 and NGC\,2403 ($\sim10-20\%$).

\subsection {HVCs and IVCs}

\begin{figure*}
\begin{center}
\includegraphics[width=\textwidth]{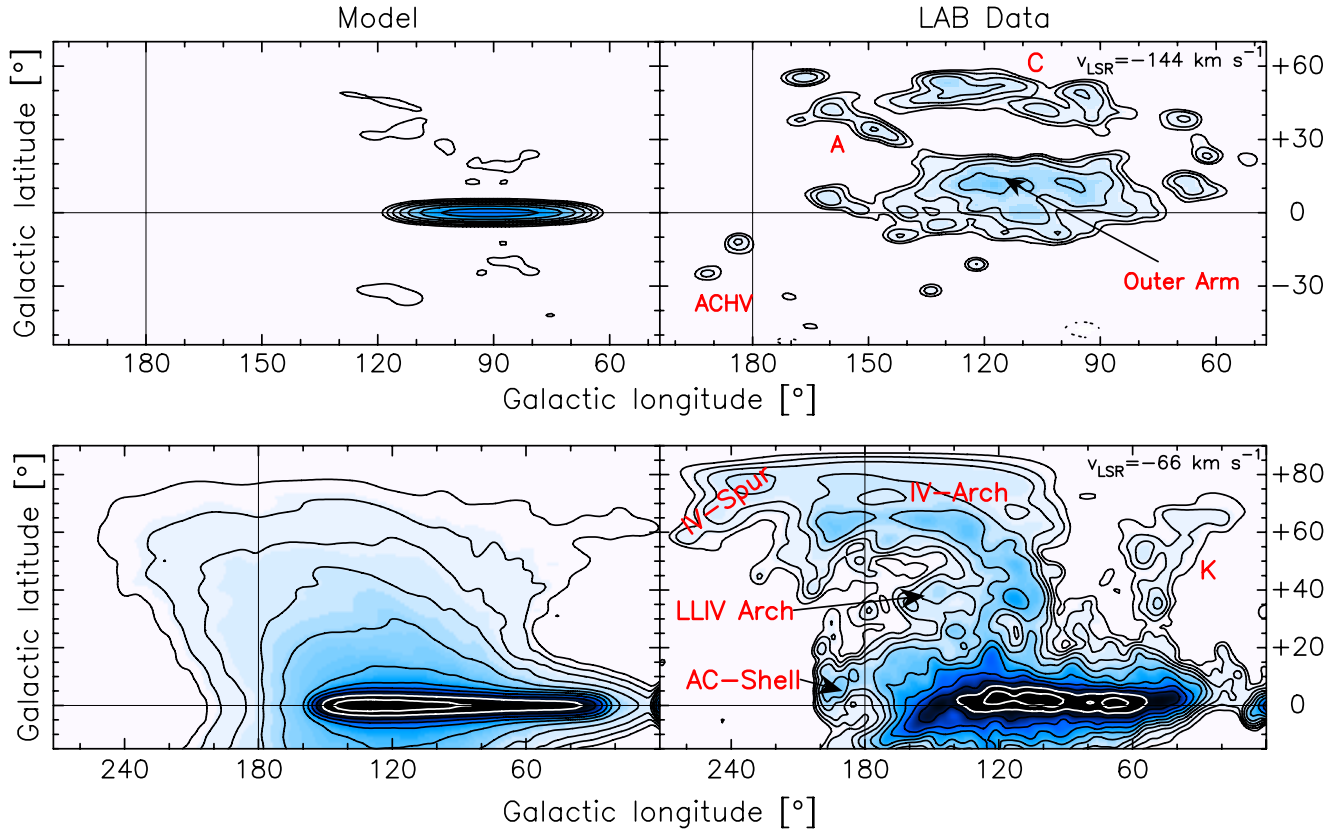}
\caption{
 Channel maps at $v_{\rm LSR}\!=\!-144\kms$ (top) and $v_{\rm
LSR}\!=\!-66\kms$ (bottom) for our best model with coronal condensation
(left panels) and the LAB data (right panels). Both model and data are
smoothed to $4\de$ resolution. Contour levels in brightness temperature range
from $0.04\K$ to $81.92\K$ scaling by a factor $2$. Some of the high and
intermediate-velocity complexes are labelled (see text).}
\label{channels}
\end{center}
\end{figure*}

At large line-of-sight velocities the \hi\ emission in the halo region of our
Galaxy is dominated by the HVCs, the emission from which lies in rather
isolated outlying regions of $(l,b,v)$ space delimited by $|v_{\rm DEV}|>90\kms$ \citep{vanWoerden+04}. The upper panels in
Fig.~\ref{channels} compare the emission predicted by our best model (left
panel) near $v_{\rm LSR}\!=\!-144\kms$ with what is actually observed. In the
observational data, three main islands of emission are apparent. At
$b>30^\circ$ we see Complexes A and C, which have no real counterparts in the
model. Similarly emission at $l\ga 180^\circ$ associated with the Anticentre
High-Velocity Cloud (ACHV) has no counterpart in the model.  It is well known
that the metallicity of most HVCs, in particular Complexes A and C is much
lower than that of gas in the disc, which strongly suggests that they come
from outside the Galaxy \citep[see][]{Wakker01}.  Moreover, assuming a
distance of $10\kpc$ for both Complexes A and C \citep{vanWoerden+04,
Thom+08} gives for these objects heights from the midplane of $6-8\kpc$.  To
produce emission so far from the plane, we would need a kick velocity $h_{\rm
v}\gtrsim150\kms$.  A model with $h_{\rm v}$ so large completely fails to
reproduce the data globally. In summary, HVCs such as Complexes A and C are
almost certainly extragalactic in origin and our model is quite correct not
to reproduce them.  The rate at which the Galaxy accretes gas from infalling
HVCs is $\sim0.2\moyr$ \citep{Sancisi08}, an order of magnitude lower than our estimate of the rate of accretion
via the fountain $+$ corona condensation mechanism.

The main body of emission in the upper right panel of Fig.~\ref{channels}
forms a large island that straddles the plane. Its summit lies at
$(l,b)=(120^\circ,15^\circ)$ and is labelled ``Outer Arm''. This feature is
thought to arise from the warp of the \hi\ disc.  Our model (left panel) does
not include a warp in the disc, so apart from two horns of very faint
emission either side of the plane, its emission is confined in a thin region
around the plane.

The lower panels of Fig.~\ref{channels} show predicted (left) and observed
(right) emission at much lower heliocentric velocities ($v_{\rm
LSR}\!\simeq\!-66\kms$) than the upper panel. At such Intermediate Velocities
(IVCs are classically defined to be at $35\kms\!<\!|v_{\rm DEV}|\!<\!90\kms$) 
the model is seen to reproduce the global structure of the data very well.
Moreover, some of the IVCs that contribute to this emission have
been shown to have a distance from the Sun of $\la3\kpc$ and disc-like
metallicities \citep{vanWoerden+04}, consistent with their being fountain
clouds.  In the lower-right panel of Fig.~\ref{channels} several IVCs are
visible (IV-Arch, IV-Spur, LLIV-Arch, AC-Shell, Complex K), mainly at
positive latitudes.  The emission of our model is smoother than the
data, but apart from that it is consistent with the existence of all the
IVCs.

The classical IVCs shown in the lower right panel of Fig.~\ref{channels} must
be just the most conspicuous members of a large population of IVCs that
collectively comprise the \hi\ halo (see also MF11). By assuming azimuthal
symmetry, our model suppresses much of the noise inherent in observing a
population of discrete clouds. In principle one could hope with a more
sophisticated model to reproduce the statistical properties of this noise,
but there is no particular merit in reproducing individual IVCs, which will
just be the chance, and ephemeral, products of individual bursts of star
formation.

If IVCs contain a mixture of metal-rich gas ejected from the disc and gas
condensed from the rather metal-poor corona, the metallicities of
these clouds should be intermediate between those of the disc and corona.
Since the fraction of accreted material is $\sim24\%$ of the whole
halo mass, the IVCs should be somewhat more metal-poor than the disc.

\section{Discussion}\label{discussion}

\subsection{Analytic approximation vs hydrodynamical simulations}
\label{compSim}

In Section \ref{SN-driven} we pointed out that our assumption of exponential
growth of the mass of the cold gas is qualitatively similar to that observed
by \citet{Marinacci+10} in their hydrodynamical simulations of cold clouds
travelling through a hot medium.  Here we show that this agreement is
quantitatively sound.  

\begin{figure}
\begin{center}
\includegraphics[width=0.48\textwidth]{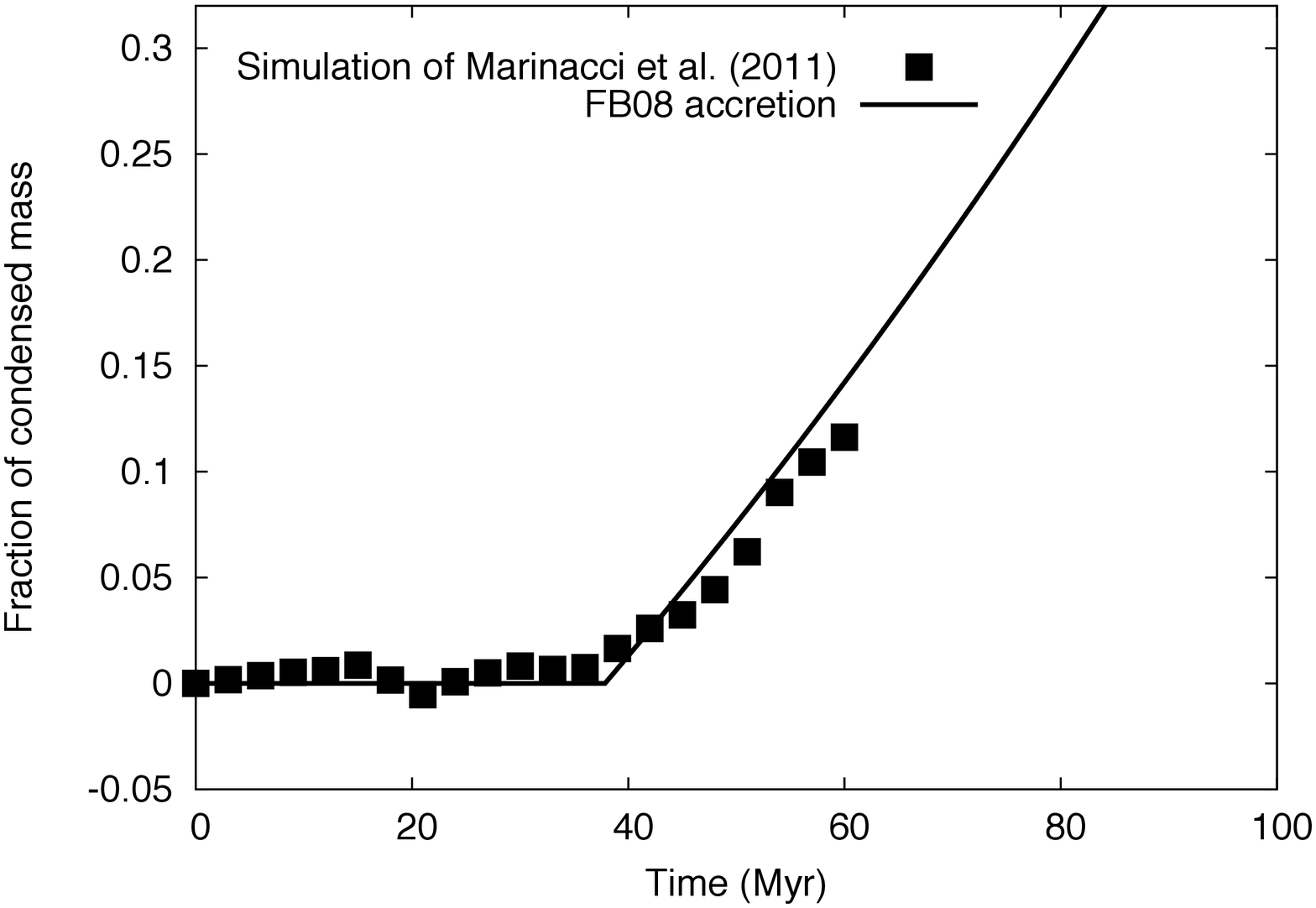}
\includegraphics[width=0.48\textwidth]{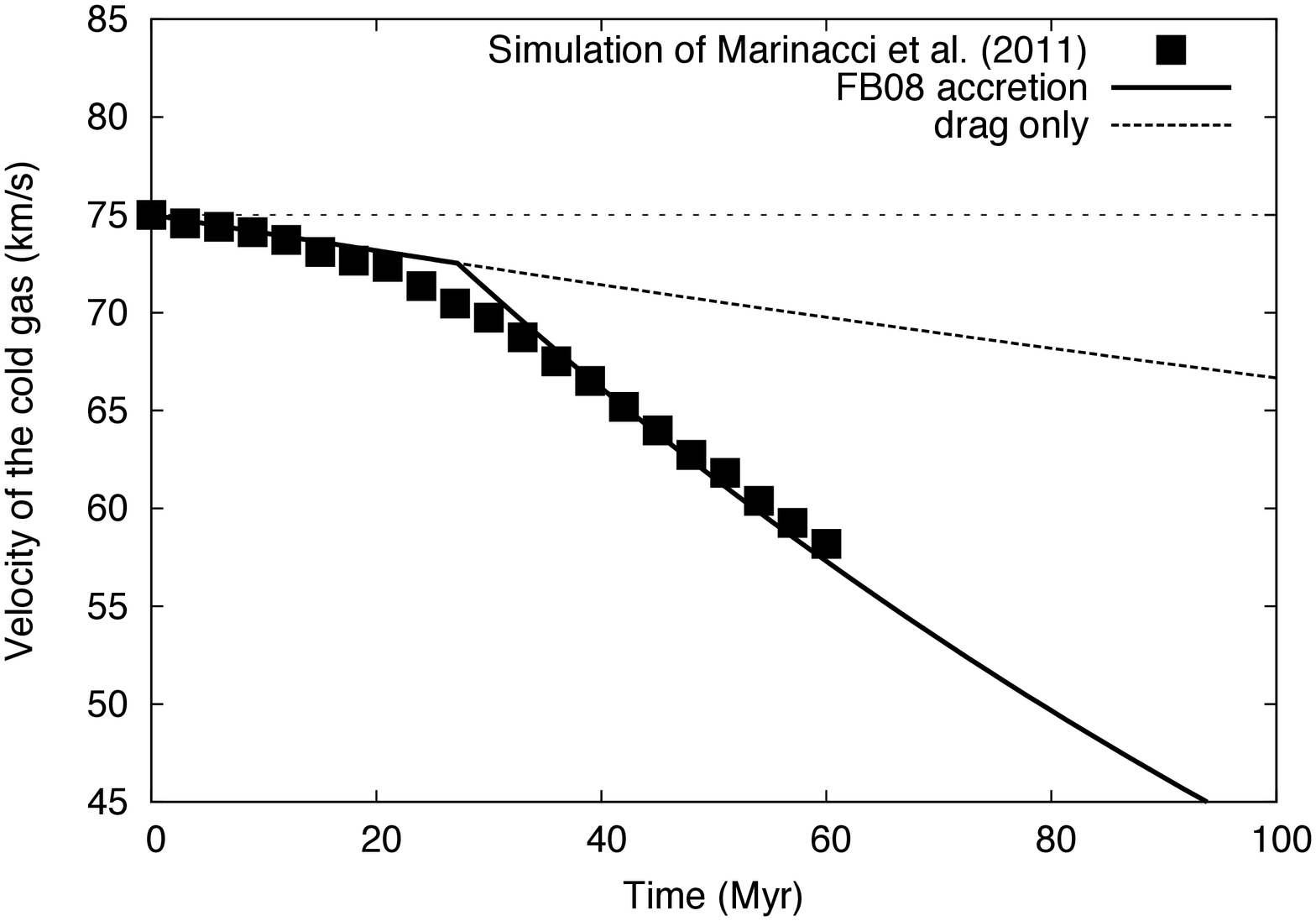}
\caption{
\emph{Top panel:}
The square points show the mass of coronal gas that condensed into the 
cloud wake as a function of time extracted from the simulations of
Marinacci et al. (2011).
\emph{Bottom panel:} The square points show the
velocity centroid of cold gas in the simulations of
Marinacci et al. (2011), the dashed curve shows the prediction in case of drag ($t_{\rm drag}=800\Myr$) without condensation.
In each panel the solid curves show the predictions of the inelastic collision+drag
recipe used in this work (equation \ref{eq:vdot}).
}
\label{fDeltaM}
\end{center}
\end{figure}

The points in the upper panel of Fig.~\ref{fDeltaM} show the mass of coronal
gas that has $T<3\times 10^4 \K$ in the simulations of \citet{Marinacci+11};
the units are fractions of the initial mass of the cloud.  
The curve shows the mass accreted by a particle that after $38\Myr$
starts to accrete according to equation (\ref{eq:mdot}) with the
growth rate set to the value, $\alpha=6.3\Gyr^{-1}$, determined by our
fits to the LAB datacube (Table~\ref{tModels}). 
The fit in Fig.~\ref{fDeltaM} between the curve and
the data points from \citet{Marinacci+10} is excellent. 
We interpret the delay by $38\Myr$ before accretion starts as the time
required for cold gas from the cloud to mix with the coronal gas plus
the time required for the coronal gas to cool to $T<3\times 10^4 \K$. 

The points in the lower panel of Fig.~\ref{fDeltaM} show the velocity
of the cold-gas centroid in the simulations of \cite{Marinacci+11}.  
The dashed line shows the evolution of the velocity if the cloud
were a rigid body that experienced hydrodynamical drag as described in
FB06 with $t_{\rm drag}\!=\!800\Myr$. 
This line fits the data from \cite{Marinacci+11} well until
$t\simeq20\Myr$. 
At later times the full thick curve shows the prediction for the
combined effects of drag and condensation, obtained from equation
(\ref{eq:vdot}) using again $\alpha=6.3\Gyr^{-1}$.
The required condensation rate is very similar than that predicted by the
simulations.

It is remarkable that detailed hydrodynamical simulations yield a value of
the accretion rate $\alpha$ that agrees so well with the value we
obtained by fitting the \hi\ datacube because the only connection
between these two determinations is the underlying physics of
turbulence. 
It is true that to obtain the agreement we did have to choose a value
for the delay between the start of a hydrodynamical simulation and the
onset of effective cooling. 
{\it Some\/} delay is inevitable because the hydrodynamical
simulations start with unrealistically spherical clouds and no wake. 
More realistic initial conditions for the dynamics of gas expelled
from the disc by a superbubble would yield a time shorter than
$30-40\Myr$ for steady accretion to become established.

The simulations of \citet{Marinacci+11} last only 60 Myr; new simulations
that also include gravity are now being carried out (Marinacci et al., in preparation).
These show that the cold gas (cloud $+$ wake) falls back to the disc in 
rather coherent structures with sizes of $1-1.5 \kpc$ and after times also larger than 60 Myr and comparable to the ballistic travel times.

While our choice of temperature threshold at $T=3\times 10^4 \K$ for
gas to become visible as \hi\ is somewhat arbitrary, experimenting
with different temperature thresholds we found that the trends of mass
and velocity are perfectly comparable with the one shown here (F.\, Marinacci, private communication).

In conclusion, the analytical treatment of an \emph{inelastic collision}
between fountain particles and ambient gas proposed by FB08 and adopted here
is quantitatively supported by the numerical simulations of
\cite{Marinacci+11}.  The system of cloud $+$ turbulent-wake accretes mass and
looses specific momentum in the way predicted by the analytical formula.  
Remarkably the rate at which this process occurs in the simulations is the same as the rate
we estimated by fitting models to the LAB survey data.

\subsection{Comparison to other galaxies}

FB06 and FB08 applied the model discussed here to two external galaxies, the
Milky Way-like NGC\,891, and the M33-like NGC\,2403.  For the former the kick
velocities necessary to reproduce the data were in the range $70-90 \kms$
depending on the shape of the potential and the extent and type of gas
accretion considered.  For NGC\,2403 the data could be reproduced with values
of the initial kick velocity from $75 \kms$ down to about $50 \kms$.  Thus
the value of $70 \kms$ that minimises the residuals in the Milky Way is fully
consistent with those needed to produced the \hi\ halos of other galaxies.
As discussed in FB06, these velocities are also consistent with
hydrodynamical simulations of superbubble expansion
\citep[e.g.][]{MacLow+89}.

The accretion parameters $\alpha$ estimated by FB08 for NGC\,891 and
NGC\,2403 are a factor $3-7$ lower than that found here for the Milky Way.
This would imply shorter timescales for the condensation of the ambient gas
in the Milky Way.  However, the value of $\alpha$ is linked to the rotational
velocity assumed for the corona where it interacts with the fountain.
FB08 assumed that the fountain gas interacted with a \emph{static} ambient
medium, which maximised the effect of the process. Had we considered
such a configuration for our corona we would have found a condensation
parameter $\alpha = 2.5\Gyr^{-1}$. In this case $t_{\rm drag}\sim300$, which implies that the drag dominates over the condensation. However, as explained in Section
\ref{SN-driven}, a static (non-rotating) corona is not a realistic
possibility.

 \section{Conclusions}\label{conclusions}

Just as in the 1980s photometry of external galaxies proved the key to making
sense of star counts within our Galaxy \citep{BahcallSoneira}, so it
is natural to turn to studies of the \hi\ distributions of external galaxies
for help in interpreting the Galaxy's \hi\ datacube. In this paper we have
investigated the extent to which the latter can be understood using a model
of the extraplanar \hi\ that emerged from observations of external galaxies
such as NGC\,891 and NGC\,2403. In this model clouds of relatively cool gas
are ejected from the plane by supernova-driven superbubbles and subsequently
orbit over the disc on near ballistic trajectories whilst weakly interacting
with the virial-temperature coronal gas through which they move. The
interaction with the coronal gas is through a combination of ram-pressure drag and
accretion of gas that cools in the cloud's wake. As a result of this
accretion, more gas returns to the plane  than left it.

In light of recent developments, we modified the model slightly before
applying it to the Galaxy. The principal modifications were (a) to allow for
rotation of the corona, and (b) to change the rule for determining the SFR
from one based on total gas density to one based on the density of molecular
gas alone. Neither modification arises from a desire to fit data for the
Galaxy better; (a) reflects an improved understanding of the hydrodynamics of
cloud-corona interaction, and (b) arises from better data for external
galaxies. A task for the future is to reanalyse the data for external
galaxies using our present model.

Our model has three adjustable parameters, the characteristic velocity of
cloud ejection $h_{\rm v}$, the specific accretion rate of clouds $\alpha$,
and a dimensionless parameter $f_{\rm ion}$, which determines where along its
trajectory the cloud's gas, which is initially photoionised, becomes visible
as \hi. To fit the data we require $h_{\rm v}\simeq70\kms$, regardless of the
assumptions we make about the values of $\alpha$ and $f_{\rm ion}$.

At high Galactic latitudes, the observations show a clear bias towards
negative velocities, and from this fact it follows that $f_{\rm ion}$ has to
be larger than zero. If we set $\alpha=0$, thus excluding
accretion from the corona, we obtain a reasonable fit to the data with
$f_{\rm ion}=1$, so clouds become neutral only when they start to move back
towards the plane. 

Permitting $\alpha$ to be non-zero yields fits to the data that are improved
in small but significant ways. In particular, with $\alpha>0$, more emission
is predicted at negative velocities at low latitudes and $l=180^\circ$. Also
less emission is predicted at negative velocities and $|l|\simeq\pm10^\circ$
and $|b|\simeq30^\circ$. Both these changes arise from an increase in the
extent of global inflow of the \hi, and they improve the fit to the data.
Moreover, with $\alpha\ne0$ the optimum value of $f_{\rm ion}$ is reduced
from unity to $0.3$, so clouds become neutral about $30$ percent of the way to
their highest point above the plane. Our best-fitting model has
$\alpha=6.3\Gyr^{-1}$, which implies that $1.6\moyr$ of hydrogen is accreted from
the corona. Including the He content, this value rises to $2.3\moyr$. This accretion rate is in excellent agreement with estimates of
the accretion rate required to sustain the Galaxy's current rate of star
formation without depleting its rather meagre stock of interstellar gas.

Unfortunately, two problems prevent us from tightly constraining the value of
$\alpha$. The first is that it is inferred from some quite subtle features in
the \hi\ datacube. The second is that the optimum value of $\alpha$ depends
on the value adopted for $v_{\rm lag}$, the equilibrium difference in the
rotation velocities of the \hi\ halo and the corona. When $v_{\rm lag}$ is
raised to $100\kms$, the optimum value of $\alpha$ decreases to
$4.0\Gyr^{-1}$, which corresponds to an accretion rate of $1.0\moyr$ ($1.4\moyr$ including the He content).

Fig.~\ref{channels} illustrates how successfully the
model simulates \hi\ emission at Intermediate Velocities -- the
simulation is probably as perfect as it can be without reproducing individual
superbubbles. By contrast, the model does not reproduce emission at High
Velocities, presumably because HVCs are extragalactic in origin.

We find remarkable agreement between the optimum values of the model's
parameters and the values found in earlier work. In particular, our value
of $h_{\rm v}$ lies within the ranges of values found for NGC\,891 and
NGC\,2403, and although our value for $\alpha$ is higher than that found for
NGC\,891 this difference can be traced to different assumptions regarding the
rotation of the corona here and in the work of FB08 on NGC\,891. Our value of
$\alpha$ does beautifully reproduce the results of a simulation by
\cite{Marinacci+11} of the hydrodynamics of cloud-corona interaction. Our
value $d v_\phi/d z\simeq-14.3\kms\kpc^{-1}$ for the vertical gradient of the
halo's mean-streaming velocity agrees perfectly with the value
($\sim-15\kms\kpc^{-1}$) determined for NGC\,891. These quantitative
agreements between studies that use either radically different methodologies
or data inspires confidence that our underlying physical picture is correct.

If we accept the fundamental soundness of the model, we can from it infer the
three-dimensional structure of the \hi\ halo, which is otherwise unknown. We
find that the halo contains $\sim3.0\times10^8\mo$ of gas, of which
$2.7\times10^8\mo$ is neutral, in agreement with the value of $3.2^{+1.0}_{-0.9}\times10^8\mo$ estimated by MF11. 
Its vertical scaleheight increases roughly linearly with radius from $\sim250\pc$ at $R=3\kpc$ to $2\kpc$ at $R=14\kpc$.
At larger radii the scaleheight falls along with the local SFR.  This \hi\
halo is neither in hydrostatic equilibrium nor strictly in circular rotation:
an analysis of the data that is nevertheless based on the assumption of
hydrostatic equilibrium and circular motion will yield a misleading structure
for the \hi\ disc, especially the flare. This structure will in turn lead to
false conclusions about the distribution of matter required to
gravitationally confine the \hi\ disc (see Sect.\ \ref{thickness}). The rate
per unit area at which the halo deposits coronal gas on the disc increases
roughly linearly from near zero at $R=3\kpc$ to a peak at $R=9\kpc$ and then
falls roughly linearly to near zero at $R=13\kpc$. Models of the chemical
evolution of the disc need the accretion profile as an input, and it will be
interesting to discover how the predictions of such models change when the
present accretion profile is adopted.


\section*{Acknowledgments}

We thank Federico Marinacci for kindly providing results from his hydrodynamical simulations. 
We also thank the referee Brant Robertson for a constructive report and helpful suggestions.
AM \& FF are supported by the PRIN-MIUR 2008SPTACC.

\bibliographystyle{mn2e}
\bibliography{mwAccr}{}

\bsp

\label{lastpage}

\end{document}